\documentclass[conference]{IEEEtran}
\IEEEoverridecommandlockouts
\usepackage{cite}
\usepackage{amsmath,amssymb,amsfonts}
\usepackage{algorithmic}
\usepackage{graphicx}
\usepackage{textcomp}
\usepackage{xcolor}
\usepackage{url}
\def\BibTeX{{\rm B\kern-.05em{\sc i\kern-.025em b}\kern-.08em
    T\kern-.1667em\lower.7ex\hbox{E}\kern-.125emX}}
    \def\undb#1{\mbox{\bf{#1}}}

\begin{document}

\title{Uncertainty-Aware Artificial Intelligence for Gear Fault Diagnosis in Motor Drives\\
\thanks{This work was supported by Innovation Fund Denmark through the project AI-Power \cite{ai-power}: Artificial Intelligence for Next-Generation Power Electronics. \textit{(Corresponding author: Subham Sahoo)}}
}

\author{
\IEEEauthorblockN{Subham Sahoo$^*$, Huai Wang and Frede Blaabjerg}
\IEEEauthorblockA{\textit{Department of Energy} \\
\textit{Aalborg University}\\
Aalborg, Denmark \\
e-mail: \{\texttt{sssa, hwa, fbl}\}@energy.aau.dk}
}

\maketitle

\begin{abstract}
This paper introduces a novel approach to quantify the uncertainties in fault diagnosis of motor drives using Bayesian neural networks (BNN). Conventional data-driven approaches used for fault diagnosis often rely on point-estimate neural networks, which merely provide deterministic outputs and fail to capture the uncertainty associated with the inference process. In contrast, BNNs offer a principled framework to model uncertainty by treating network weights as probability distributions rather than fixed values. It offers several advantages: (a) improved robustness to noisy data, (b) enhanced interpretability of model predictions, and (c) the ability to quantify uncertainty in the decision-making processes. To test the robustness of the proposed BNN, it has been tested under a conservative dataset of gear fault data from an experimental prototype of three fault types at first, and is then incrementally trained on new fault classes and datasets to explore its uncertainty quantification features and model interpretability under noisy data and unseen fault scenarios.
\end{abstract}


\section{Introduction}
Models developed using deep learning are widely used in all types of inference and decision making in the field of power electronics \cite{shuai}. In other words, it is becoming increasingly important to assess the reliability and effectiveness of artificial intelligence (AI) models before putting them into practice. This is because the predictions of AI are usually affected by noise and model output errors, leading to unexplainable results \cite{sahoo}. 
These uncertainties arise when a mismatch between the testing and training data is encountered. Although these uncertainties can have a significant impact on the trained AI model's estimation capabilities, it is difficult to compensate for the uncertainties arising out of model knowledge uncertainty. As a result, such uncertainties in data and models can be segmented into two categories, namely \textit{aleatoric} and \textit{epistemic} uncertainty.
\begin{figure*}[h]\centering
	\includegraphics[width=0.85\linewidth]{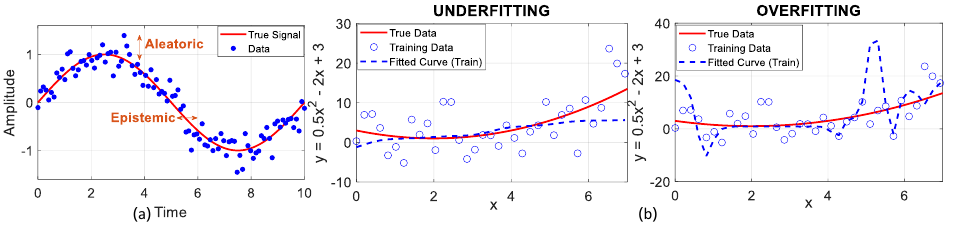}
	\caption{(a) Schematic view of the categorical differences between aleatoric and epistemic uncertainty -- the former is aimed at noisy diverging data and the latter focused on missing information, (b) Overfitting issue caused by NNs due to training over limited data -- as the training data corresponding to the polynomial $y = 0.5x^2-2x+3$ is collected aimed at regressing over the true data, overfitting over minimal points can cause a large deviation from the actual model.}
    \label{FIG_1}
    \vspace{-6 pt}
\end{figure*}
\subsection{Classification of uncertainties in AI}
By definition, statistical inconsistencies in data leading to prediction uncertainty by an AI model is called as \textit{aleatoric} uncertainty (commonly referred as data uncertainty).
This type of uncertainty is an inherent property of inconsistent data distribution, which ultimately becomes a barrier in distinguishing between overlapping data groups. 

In contrast, \textit{epistemic} uncertainty (commonly referred as model uncertainty) occurs due to inadequate knowledge of the model. Even in scenarios when the data is sufficient, their valuation can still be deemed as information-poor from a contextual data collection perspective. In such cases, AI-based methods are usually referred to characterize the emergent features of the data. However, since the data required for developing AI-based methods can be rather incomplete, noisy, discordant, the predictions are not always accurate. This aspect has been accounted in Fig. \ref{FIG_2}(a) performed on a sinusoidal signal, and corresponding data is extracted to map the \textit{true signal}. Based on our definitions above, it can be seen that some data do not really characterize themselves close to the sinusoidal variations and naturally increase the aleatoric uncertainty (around t = 3.5 sec), whereas the epistemic uncertainty is seen around t = [5, 6] sec, where missing data doesn't transcend to the actual model information.
\subsection{Key reasons behind uncertainties in AI predictions}
{\textbf{Limited data:}}
Going beyond limited training to have an highly accurate ensemble model, there are many practical scenarios which does not invoke more data because of the risks. One such example for power electronic applications is fault based scenarios. Since fault data is limited and risky to be emulated in the lab, NNs can easily provide overconfident decisions when trained over limited set of data. This has been clearly illustrated in Fig. \ref{FIG_1}(b), where overfitting over less data can cause large deviations from the actual polynomial model trajectory (in red). 

{\textbf{Unseen data/scenarios:}} 
The state-of-the art methods rely on the training data with a strong assumption that it is well connected with out-of-distribution (OOD) data. However, it can be a very strong assumption specifically for research in fault diagnosis, where the nature, type and properties of a new fault can significantly vary from the past training data. Basically, the state-of-the-art methods rely on a \textit{frequentist} deep learning approach, that represents emergent behavior of data merely as deterministic values to deliver a point-estimate prediction. These predictions, in turn, can often be over-confident and project a false representation about the system due to unseen data.

\begin{figure}[h]\centering
	\includegraphics[width=0.8\linewidth]{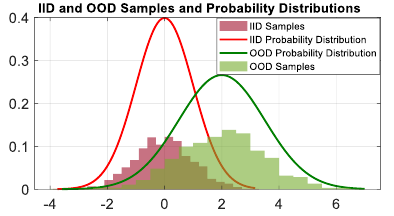}
	\caption{Out of distribution (OOD) samples correspond to the unseen data/conditions, that ultimately  aggravates the uncertainty in deep learning predictions.}
    \label{FIG_2}
    \vspace{-6 pt}
\end{figure}

This gap has been illustrated in Fig. \ref{FIG_2}, which highlights the training data as the independent identically distributed variables being the foundation behind the learning policy of neural networks (NN). One of the primary pre-requisite behind training a NN is that the training and testing data must follow independently identically distribution (IID) pattern, as shown in Fig. \ref{FIG_2}. However, this is not intuitive in real-world applications where the data on which a NN is trained and finally tested might differ significantly. This leads to an out-of-distribution (OOD) problem that often leads to unreliable decisions for unforeseen data.
As evident, the OOD or testing samples in Fig. \ref{FIG_2} accounting for an unseen condition/scenario will be seen as an adversarial condition by the NN, and can easily lead to under/overconfident decisions merely based on limited statistical insights. 
\subsection{Literature survey}
As unpredictable faults in general can be, mechanical faults in machines pose a significant risk which necessitates intelligent fault diagnostic methods and health monitoring tools \cite{zio,hu}. The automation of such diagnostic procedures has been made more intelligent by using AI to set an alarm for dynamic contingent conditions. Although model-based signal processing approaches, such as wavelet analysis, empirical mode decomposition (EMD) have been vital in extracting faulty signatures \cite{5,6}, they are not directly compatible with advanced and intelligent data-driven methods. Furthermore, they only provide a low automation degree.

Since fault diagnosis is particularly a feature engineering task, data-driven approaches in the form of AI tools are well-equipped and pronounced to determine any fault signatures and can seamlessly update the library of faults simply by introduction of new faulty data \cite{9}. Many deep learning models, such as stacked encoders (SAE), convolutional neural networks (CNN) have been utilized for fault diagnosis before \cite{12}-\cite{14}. To compensate for the lack of interpretability in these models, many grey-box models have been developed to bridge the gap between domain knowledge and data-driven predictions \cite{19}-\cite{20}. Since these are point-estimate algorithms that only infer with deterministic values, new probabilistic data-driven approaches in the form of Bayesian neural networks \cite{bnn} have been exploited for fault diagnosis of mechanical devices. However, they still lack a principled way of testing and model structuring -- without providing any direct guideline on an optimal selection of the number of layers, neurons or variational inferences.

\subsection{Main contributions}
Based on the issues discussed above, this paper exploits uncertainty-aware Bayesian neural networks (BNN) and its effectiveness for fault diagnosis of gear box faults emulated on a fault simulator. A preliminary case study has been thoroughly covered in \cite{agni} using point-estimate neural networks, which only generalizes the decisions based on accuracy without any explainability measures. Some of the key features of BNN that favor in minimizing the uncertainty of predictions for power electronics are:
\begin{itemize}
\item \textbf{Probabilistic outputs}: BNNs provide probabilistic outputs rather than point estimates. This means that instead of predicting a single value, they provide a distribution over possible outcomes, capturing uncertainty in predictions.
\item \textbf{Uncertainty estimation}: They offer a principled way to estimate uncertainty associated with predictions. This uncertainty can be categorized into aleatoric uncertainty (inherent randomness or noise in the data) and epistemic uncertainty (uncertainty due to limited data or model uncertainty).
\item \textbf{Bayesian inference}: Bayesian neural networks use Bayesian inference techniques to learn model parameters. Instead of finding a single set of parameters that maximize a likelihood function (as in traditional neural networks), they learn a distribution over parameters given the data, incorporating prior knowledge and updating beliefs based on evidence.
\end{itemize}
\section{System Preliminaries \& Problem Statement}
\begin{figure}[h]
\centering
\includegraphics[width=0.22\textwidth]{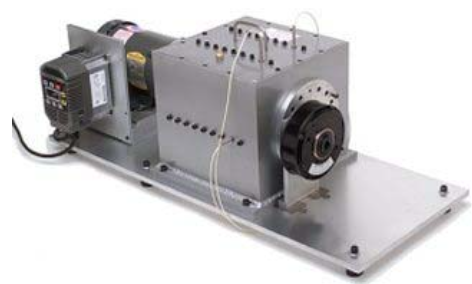}
\caption{Gearbox Dynamics Simulator used for collecting fault data. Detailed setup specifications can be found in \cite{agni}.} \label{fig6}
\end{figure}

To emulate such faults, SpectraQuest’s Gearbox Dynamics Simulator (GDS), shown in Fig. \ref{fig6}, is used to simulate industrial gearboxes in both educational and experimental applications. It is highly precise and comprehensive, featuring a two-stage parallel shaft gearbox with rolling bearings and a magnetic brake system. This intricate design provides an ideal platform for advanced insights into the multifaceted dynamics and acoustic behavior of gearboxes. 
More details on this setup can be found in \cite{agni}.

\textbf{Acquisition:} The GDS has five potential fully developed faults on the spur and helical gears. Gear faults typically manifest as cracks on the gear or wear and tear of the gear teeth.

One of the methods employed in various reports is gear fault detection using vibration analysis. For vibration analysis, gearboxes are typically mounted with an acceleration sensor on the gearbox housing. A healthy gearbox theoretically has the dominant vibration mode in the axial direction; the vibration frequency is also called the gear-mesh frequency. More details on the gear mesh spectrum can be found in \cite{agni}.
\subsection{Data \& Setup Specifications}
The setup in Fig. \ref{fig6} consists of the following actuators and sensors:
\subsubsection{Actuators}
\begin{itemize}
    \item {Motor (Reconfigurable)} -- This unit is multiple copies of a 3-phase motor which is configurable to the following states:
\begin{enumerate}
\item No fault (3HP)
\item rotor unbalance fault (1HP)
\item rotor misalignment fault (1HP)
\item bowed rotor fault(1HP)
\item broken rotor fault (1HP)
\item stator winding fault (1HP)
\item voltage unbalance and single phasing (1HP)
\end{enumerate}
\item {Gearbox (Reconfigurable)} -- The experimental set-up houses a gearbox, which can be easily swapped. They can be reconfigured to the following faults:
\begin{enumerate}
\item Missing tooth gear 
\item Chipped tooth gear 
\item Root crack gear 
\item Surface wear gear 
\item Eccentricity 
\end{enumerate}
\item {Brake (Reconfigurable load)} -- A programmable magnetic brake (24.8567 N-m) that emulates gearbox loading.
\end{itemize}
\subsubsection{Sensors}
Two types of sensors were used in this project, intrinsic and extrinsic:
\begin{itemize}
\item \underline{\textit{Intrinsic Sensors}}: The setup in Fig. \ref{fig6} was modified with a Danfoss VLT Drive FC-103 to provide the following intrinsic parameters as outputs:
\begin{enumerate}
    \item \text{Speed}
    \item \text{Motor Torque}
    \item \text{DC-link voltage} 
    \item \text{Reactive stator current}
\item Active stator current
\item Motor power
\end{enumerate}
sampled at a frequency of 5 kHz. 

\item \underline{\textit{Extrinsic Sensors}}: The gearbox was outfitted with a pair of orthogonally aligned analogue accelerometers ADXL1001.
\end{itemize}
\begin{figure}[h]
\centering
\includegraphics[width=0.3\textwidth]{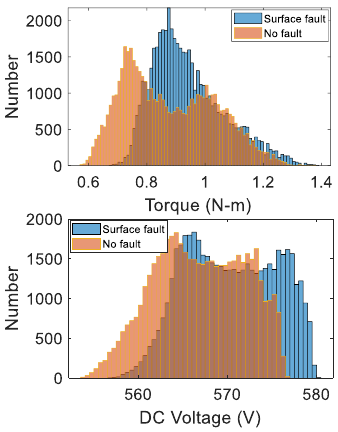}
\caption{Fault signatures for the same loading profile -- surface fault vs. no fault. The overlapping region for both torque as well as DC voltages can lead to over-confident decisions from AI models due to conventional point-estimate deterministic learning approaches.} \label{fig7}
\end{figure}
\subsection{Challenges with current diagnostic approaches}
Considering a simple comparative example of extrinsic sensor data on surface faults against the datasets representing no fault (see Fig. \ref{fig7}), it can be seen that there is no visible statistical difference for torque or DC voltage profile, since the distribution shifts on the application of load. This could potentially be due to the lack of further application of filters or little understanding of how the parameters interact with each other. As explained before, this affects the decisions taken by AI, which is primarily driven by these statistical attributes.
\section{Uncertainty-Aware AI}
Despite of promising applications offered by deep learning (DL) methods for power electronics and motor drives, the lack of interpretability and uncertainty quantification in their decisions is a significant barrier with their implementation. Hence, we propose a customized uncertainty-aware Bayesian neural network (BNN) for quantifying the uncertainty in fault diagnosis of the gear boxes in Fig. 3. Since the diagnosis is performed on extrinsic signals, the variational inference is tweaked with model information to achieve high generaltion. 
\subsection{Theory}
Before discussing Bayesian modeling principles, let us start with the preliminaries of a simple feed forward neural network (NN) to understand uncertainty modeling in
detail.

Consider a preliminary structure of a NN \cite{nn} having multiple layers with $\undb{x}$ be a $D$-dimensional input vector, bias $b$ and a linear mapping function $W_1$ for its transformation into a vector of $Q$ elements, given by $W_1$\undb{x} + $b$. On top, activation function $\sigma$(.) to smoothen the output of hidden layers. As a result, a multi-layer inference with another cascaded linear function $W_2$ can be given by:
\begin{equation}
    \hat{y} = \sigma(W_1\undb{x}+b)W_2.
\end{equation}

Since fault diagnosis is primarily a classification task relying on feature engineering to determine intrinsic faulty signatures, we exploit a probabilistic approach to determine the possibility of \undb{x} exclusively belonging to a certain class \{1,...,$C$\}. Finally, the score is obtained by computing the model output $\hat{y}$ with a softmax function $\hat{p}_d$ = exp($\hat{y}_d$)/($\sum_{d'}$ exp($\hat{y}_d$)). Hence, the softmax loss is calculated using:
\begin{equation}
E^{W_1,W_2,b}(X,Y) = -\frac{1}{N}\sum_{i=1}^N \text{log}(\hat{p}_{i,c_i}).
\end{equation}
where, $X$ = \{$\undb{x}_1$,...,$\undb{x}_N$\} and $Y$ = \{$y_1$,...,$y_N$\} are the model's input and output vectors, respectively.


\subsection{Uncertainty Modeling}
Predictive uncertainty (PU) consists of two parts: (i) epistemic uncertainty (EU), and (ii) aleatoric uncertainty (AU), and can be represented as their sum:
\begin{equation}
    PU = EU + AU.
\end{equation}
Let $D_{tr}$ = \{$X$, $Y$\} = \{($\undb{x}_i$, $y_i$)\}$_{i=1}^N$ denote a training dataset with inputs $\undb{x}_i \in \mathbb{R}^D$ and outputs $y_i \in \{1,...,C\}$, where $C$ represents the number of classes. Given both types of uncertainties in (3), the objective is to optimize the parameters $\omega$ in $y$ = $f^{\omega}$(\undb{x}) and obtain the desired output. To achieve this using a probabilistic approach, {Bayesian methodologies} define a model likelihood, $p(y | \undb{x},\omega)$. For classification, \textit{softmax} likelihood will be obtained using:
\begin{equation}
    p(y = c|\undb{x},\omega) = \frac{\text{exp}(f_c^{\omega}(\undb{x}))}{\sum_{c'}\text{exp}(f_{c'}^{\omega}(\undb{x}))}.
\end{equation}


For a given test sample $\undb{x}^*$, the probability of identifying a class label with regard to $p(\omega|X, Y)$ can be predicted using:
\begin{equation}
    p(y^*|\undb{x}^*, X, Y) = \int p(y^*|\undb{x}^*, \omega)p(\omega|X, Y) d\omega
\end{equation}
However, $p(\omega|X,Y)$ in (5) cannot be computed analytically. Hence, we exploit variational inference algorithms \cite{hands} to approximate the variational parameters, i.e., $q_{\theta}(\omega)$. Hence, the rationale behind using variational inference algorithms is to approximate a distribution for each neuron, such that it is close to the posterior distribution obtained by the model.

It is worthy notifying that the Bayesian inferencing in statistics is strongly correlated  with the frequentist paradigm, primarily used for hypothesis testing of training data. In simple terms, it can be summarized by the following:
\begin{itemize}
\item probability is a measure of trust in the occurrence of events, rather than a limit in the frequency of occurrence.
\item prior inferences influence posterior output, as stipulated in the Bayes’ theorem, which can be mathematically stated as:
\begin{equation}
P(H|D) = \frac{P(D|H)P(H)}{P(D)} = \frac{\text{Likelihood}*\text{Prior}}{\text{Evidence}}
\end{equation}
\end{itemize}
It is worth notifying that $H$ and $D$ in (6) are considered as the sets of outcomes. The Bayesian inferencing rules considers $H$ to be a hypothesis about which one holds some prior belief, and $D$ to be some data that will update one’s belief about $H$. Catering back to the fault diagnosis example in this paper, $H$ would imply on faulty signatures collected from iteratively processed feature engineering on different conditions. Whereas, $D$ would be the dataset that confirms this hypothesis. Following this logic over multiple layers in BNN, the  probability distribution $P(D|H)$ in (6) is commonly termed as \textit{likelihood}, since it is basically an evaluation stage for the hypothesis over the given dataset. It encodes the aleatoric uncertainty in the model. 
Finally, $P(H|D)$ is termed as the \textit{posterior}, which ultimately encodes the epistemic uncertainty.

\subsection{Inferential Algorithm}
\begin{figure}[h]
\centering
\includegraphics[width=0.45\textwidth]{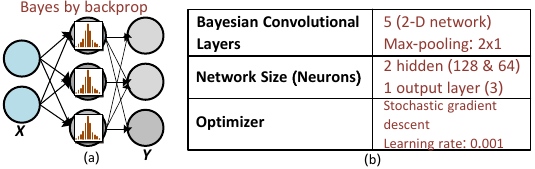}
\caption{(a) Probabilistic assignment of neuronal weights to formalize a variational inference approach, (b) Model specifications of the designed bayesian neural network (BNN).} \label{fig7}
\end{figure}
\begin{figure*}[thb]
\centering
\includegraphics[width=1.7\columnwidth]{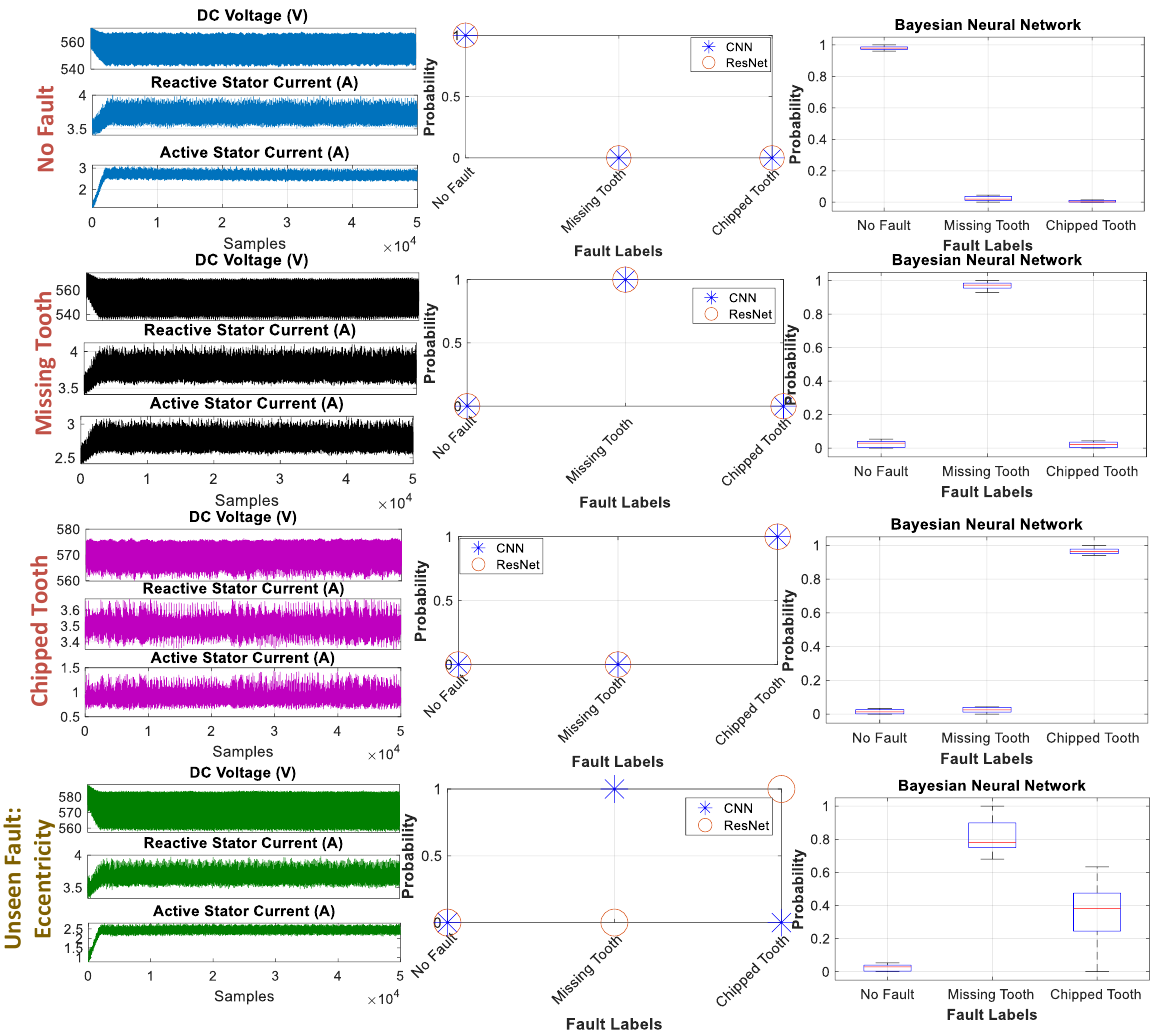}
\caption{Diagnosis of the testing samples of seen domain and unseen fault with benchmarking models and probabilistic Bayesian NN: For the extrinsic measurements considered for the three fault labels \{\texttt{No Fault, Missing Tooth, Chipped Tooth}\}, the diagnostic results of benchmarking models trained in the seen domain are fairly accurate for all the three candidate classification algorithms. However, a big discrepancy arises for an unseen fault where the uncertainty in BNN predictions rises to an alarming value.} \label{fig18}
\end{figure*}
In this paper, we will exploit the variational posterior using the well-known Bayes-by-backprop (BBB) method \cite{bbb}. Basically, this algorithm introduces a random variable $\epsilon$ having a given probability density and a deterministic transform $t(\theta, \epsilon)$, such that the weights of the BNN $w$ can be equivalent to $t(\theta, \epsilon)$. This formalizes a variational inference approach to carry out a probabilistic evaluation of uncertainty introduced by AI models, as shown in Fig. 5(a).

The key principled mechanism behind BBB method is that the random variable $\epsilon$ can define the variational distribution by assigning different values to each set-point in the distribution function, and consequently shape the weights $w$ as an indirect deterministic transformation of $\epsilon$. As a result, instead of a point-estimate inference for each condition in a conventional NN, BNN offers a range of weights that can be used to check the likelihood from input to the output. Indeed, by writing $w$ as $w = t(\theta, \epsilon)$, in place of evaluating:
\begin{eqnarray}
    \frac{\partial}{\partial \theta}\mathcal{E}_{q(w | \theta)} [f(w,\theta)] = \mathcal{E}_q(\epsilon) [\frac{\partial}{\partial \theta}f(t(\theta, \epsilon), \theta)]\\
    = \mathcal{E}_{q(\epsilon)}[\frac{\partial f(w, \theta)}{\partial w}\frac{\partial w}{\partial \theta}+\frac{\partial f(w, \theta)}{\partial \theta}]
\end{eqnarray}
we guarantee that the convergence over probabilistic variational posteriors can be achieved. The reason behind using backpropagation using Bayesian paradigm over commonly used Markov Chain Monte Carlo (MCMC) approach is due to the faster convergence of predictions and high interpretability.
\section{Performance Evaluation}
In this section, we will evaluate the performance of the modeled BNN from two different perspectives: (a) its capability with handling seen data and its accuracy estimates, (b) its capability with handling unseen conditions and data. This will provide a multi-faceted leap not only by quantifying uncertainties in predictions by AI models, but also quantifying its confidence levels alongside its predictions as another dimension of decision making. The model specifications of the BNN used in the following studies can be found in Fig. 5(b). The BNN was trained with the datasets of fault labels: \texttt{No Fault, Missing Tooth, Chipped Tooth}. As a result, the rest of the fault labels will be considered as an \textit{unseen} condition by the BNN.

To provide a comparative evaluation, we have considered three candidates for data-driven classification: BNN, Convolutional neural network (CNN), and ResNet, trained over the seen datasets. Since CNN and ResNet are deterministic tools, their output only indicates a single point-estimate of the probability of each fault, whereas Bayesian neural network provides a range of predictive uncertainty for the seen datasets, based on the hypothesis $H$ and \textit{likelihood} calculated over multiple BNN layers. However, when all the three algorithms are tested under the unseen \texttt{Eccentric Fault} in Fig. \ref{fig18}, their predictions differ by a very high degree, as CNN and ResNet provide conflicting probabilistic outputs. However, the same can also be implied for BNNs, where the confidence level behind the probabilistic estimates is low for all the three seen fault conditions. In any case, this visualization offered by BNN could be seen as a powerful prospect in differentiating that the new dataset doesn't fall into any of the fault category and require further investigations.

\begin{figure}[thb]
\centering
\includegraphics[width=\columnwidth]{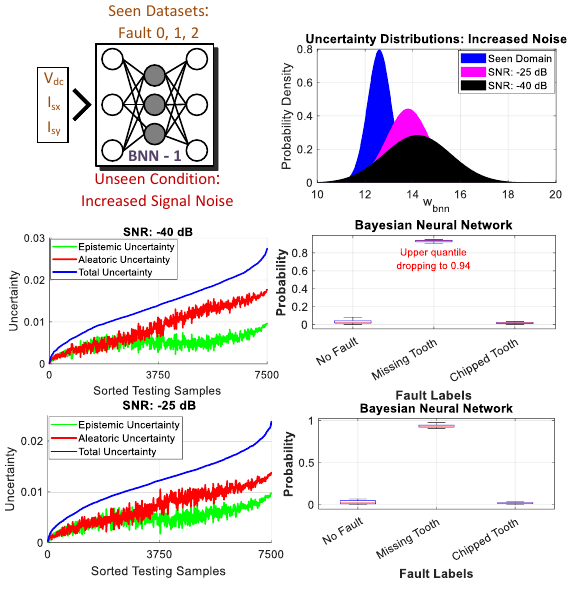}
\caption{Uncertainty estimation and decomposition using BNN for the testing samples of seen domain and different noisy environments:  the upper quantile of the predictions (refer to the box plots) drop to 0.94. With increase in noise, the aleatoric uncertainty dominate over the epistemic uncertainty.} \label{fig20}
\end{figure}
\begin{figure}[thb]
\centering
\includegraphics[width=\columnwidth]{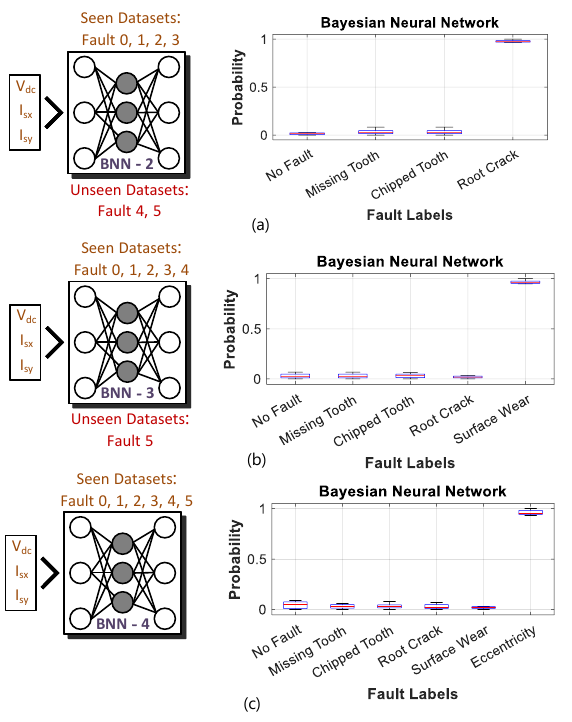}
\caption{Performance evaluation of updated seen datasets by transforming BNN into: (a) BNN--2 by adding fault label 3 into the training database onto BNN, (b) BNN--3 by adding fault label 4 into the training database onto BNN--2, (c) BNN--4 by adding fault label 5 into the training database onto BNN--3. } \label{fig22}
\end{figure}
In Fig. \ref{fig20}, the modeled BNN is tested against different noise levels, which clearly indicates that the total uncertainty increases for higher noise level. When the SNR is less than -25 dB which is indicative of a high noise profile, the proposed BNN outputs a significant rise in the aleatoric uncertainty. This is well aligned with the fact that noise, in particular, increases the level of data based uncertainties.
As a result, the change in aleatoric uncertainty becomes highly prominent over epistemic uncertainty with a decrease in SNR.

BNN trained only with the seen datasets on fault label {\texttt{No Fault, Missing Tooth, Chipped Tooth}\} is reconfigured to include the new fault cases into the seen dataset. Hence, we incorporate the new fault datasets in a step-wise manner so that three new BNN designs can be made. As shown in Fig. \ref{fig22}(a), BNN--2 is updated with fault 3 in its seen environment, which automatically minimizes the diagnostic uncertainty and accurately identifies fault 3 (root crack fault) into the correct category, as opposed to the wrong predictions made initially in Fig. \ref{fig18}. On the other hand, BNN--2 trained only with the seen datasets on fault label {\texttt{No Fault, Missing Tooth, Chipped Tooth, Root Crack}\} is reconfigured to include the new fault cases into the seen dataset. We then incorporate the new fault datasets in a step-wise manner so that three new BNN designs can be made. As shown in Fig. \ref{fig22}(b), BNN--3 is updated with fault 4 in its seen environment, which automatically minimizes the diagnostic uncertainty and accurately identifies \texttt{Surface Wear} into the correct fault category. Finally, BNN--3 trained only with the seen datasets on fault label {\texttt{No Fault, Missing Tooth, Chipped Tooth, Root Crack, Surface Wear}\} is reconfigured to include the new fault cases into the seen dataset. We then incorporate the new fault datasets in a step-wise manner so that three new BNN designs can be made. As shown in Fig. \ref{fig22}(c), BNN--4 is updated with fault 5 in its seen environment, which automatically minimizes the diagnostic uncertainty and accurately identifies \texttt{Eccentricity fault} into the correct fault category.

In this way, the unseen conditions can be gradually augmented to improve an accurate diagnosis model that not only provide reliable predictions, but also highlight the confidence interval behind each prediction.
\section{Conclusion}
\begin{figure}[thb]
\centering
\includegraphics[width=\columnwidth]{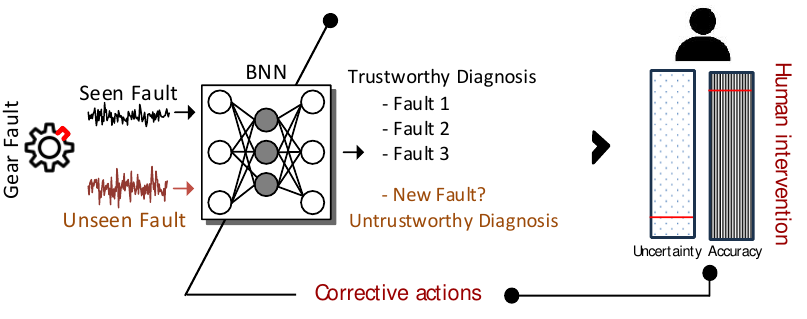}
\caption{Bayesian neural networks is a promising solution, that offers a multi-dimensional decision and allows a qualitative assessment of its predictions.}\label{fig23}
\end{figure}
In conclusion, this paper delves into the application of uncertainty-aware AI algorithms for fault diagnosis of gear faults, with a particular focus on Bayesian neural networks (BNNs). Through the utilization of BNNs, we have effectively quantified uncertainties in predictions, providing probabilistic outputs that offer valuable insights into the reliability of diagnostic assessments. We have carried out rigorous test cases by considering different subsets of faults as the training data and identifying reasonable answers from BNN for the unseen faults in a structural way. The effect of noise variance, model parameters and unseen data has been covered in detail with key results based on theoretical foundations. This provides us with a formidable framework for a multi-dimensional decision making process (see Fig. \ref{fig23}) that requires human intervention before finalizing the data-driven algorithm for fault diagnosis of power electronics. This enhances a qualitative assessment of predictions from AI that often goes overlooked due to high accuracy as the sole decision metric. As a follow-up, the decision making process will ascertain corrective actions across the input stage such that further data analysis and investigations can be carried out. 

As a result, this work provides a significant leap towards trustworthy machine learning in power electronics by quantifying the uncertainty in the predictions of AI, as a direct measure of either data-driven or model-driven uncertainties.




\vspace{12pt}

\end{document}